\documentclass[aps,pre,twocolumn,superscriptaddress,showpacs]{revtex4-2}
\usepackage{graphicx}
\usepackage{array}
\usepackage{amsfonts,amssymb,amsmath,mathrsfs,multirow,rotate}
\usepackage{booktabs}
\usepackage{soul}
\usepackage{threeparttable}
\usepackage{multirow}
\usepackage{epsfig}
\usepackage{threeparttable}
\usepackage{chngpage}
\usepackage{latexsym}
\usepackage{dcolumn}
\usepackage{graphics,graphicx,bm,fleqn,epic,eepic,float}
\usepackage{verbatim}
\usepackage{color}
\usepackage{xr}
\usepackage{float}
\usepackage{listings} % For including code in appendix
\usepackage{placeins} % For putting plots in the desired location
\usepackage{tikz}
\usepackage{url}
\usepackage{natbib}

\begin{document}

\title{Machine-learning prediction of tipping with applications to the Atlantic Meridional Overturning Circulation}

\date{\today}

\author{Shirin Panahi}
\affiliation{School of Electrical, Computer, and Energy Engineering, Arizona State University, Tempe, AZ 85287, USA}

\author{Ling-Wei Kong}
\affiliation{School of Electrical, Computer, and Energy Engineering, Arizona State University, Tempe, AZ 85287, USA}

\author{Mohammadamin Moradi}
\affiliation{School of Electrical, Computer, and Energy Engineering, Arizona State University, Tempe, AZ 85287, USA}

\author{Zheng-Meng Zhai}
\affiliation{School of Electrical, Computer, and Energy Engineering, Arizona State University, Tempe, AZ 85287, USA}

\author{Bryan Glaz}
\affiliation{Army Research Directorate, DEVCOM Army Research Laboratory, 2800 Powder Mill Road, Adelphi, MD 20783-1138, USA}

\author{Mulugeta Haile}
\affiliation{Army Research Directorate, DEVCOM Army Research Laboratory, 6340 Rodman Road, Aberdeen Proving Ground, MD 21005-5069, USA}

\author{Ying-Cheng Lai} \email{Ying-Cheng.Lai@asu.edu}
\affiliation{School of Electrical, Computer, and Energy Engineering, Arizona State University, Tempe, AZ 85287, USA}
\affiliation{Department of Physics, Arizona State University, Tempe, Arizona 85287, USA}

\begin{abstract}

% 150 words
Anticipating a tipping point, a transition from one stable steady state to another, is a problem of broad relevance due to the ubiquity of the phenomenon in diverse fields. The steady-state nature of the dynamics about a tipping point makes its prediction significantly more challenging than predicting other types of critical transitions from oscillatory or chaotic dynamics. Exploiting the benefits of noise, we develop a general data-driven and machine-learning approach to predicting potential future tipping in nonautonomous dynamical systems and validate the framework using examples from different fields. As an application, we address the problem of predicting the potential collapse of the Atlantic Meridional Overturning Circulation (AMOC), possibly driven by climate-induced changes in the freshwater input to the North Atlantic. Our predictions based on synthetic and currently available empirical data place a potential collapse window spanning from 2040 to 2065, in consistency with the results in the current literature.

\end{abstract}

\maketitle

\section*{Introduction} \label{sec:intro}

A tipping point in nonlinear and complex dynamical systems is referred to as a
transition from one stable steady state supporting the normal functioning of the
system to another that can often be catastrophic and corresponds to system
collapse~\cite{Scheffer:2004}. This can happen as a system parameter passes 
through a critical point. For example, in ecosystems, before tipping the system is 
in a survival state with healthy species populations, while the state after the 
tipping is associated with extinction~\cite{Scheffer:2004,Schefferetal:2009,Scheffer:2010,WH:2010,DG:2010,CLLLA:2012,BH:2012,DVKG:2012}. 
In the past decade, tipping point in ecosystems has been extensively 
studied~\cite{Scheffer:2004,Schefferetal:2009,Scheffer:2010,WH:2010,DJ:2010,CLLLA:2012,BH:2012,DVKG:2012,ashwin2012tipping,LLDvanNS:2012,Barnoskyetal:2012,BH:2013,TC:2014,LNSB:2014,LCJL:2015,GTZB:2015,JHSLGHL:2018,YLTLZCX:2018,JHL:2019,scheffer2020critical,MJGL:2020,MLG:2020,MG:2021,MLG:2022,OW:2020}. 
The phenomenon of tipping can also arise in other contexts such as epidemic 
outbreak~\cite{trefois2015critical}, a sudden transition from normal to depressed mood 
in bipolar patients~\cite{bayani2017critical}, alterations in the stability of the 
Amazon rain forest~\cite{nobre2009tipping}, an increase in the carbon emission from 
Boreal permafrost~\cite{miner2022permafrost}, and the melting of Arctic sea 
ice~\cite{wadhams2012arctic}. A likely scenario by which a tipping point can occur is
when a parameter of the system varies with time - nonautonomous dynamical systems. 
Suppose the system is in a normal functioning state at the present. Due to the 
parameter change, at a certain time in the future a critical point will be crossed, 
leading to a catastrophic tipping. The global climate change is causing ecosystems 
and climate systems of different scales to become such nonautonomous dynamical systems 
with the increasing risk of tipping. Articulating effective methods to reliably 
anticipate tipping is an urgent problem with broad implications and applications.

In this paper, we develop a reservoir-computing framework tailored to anticipating 
tipping in nonautonomous dynamical systems and demonstrate its predictive power 
using examples from different fields. A particular application that provided the 
main motivation for our work is predicting the possible collapse of the Atlantic 
Meridional Overturning 
Circulation (AMOC)~\cite{buckley2016observations,lohmann2021risk,jackson2022evolution} 
that supports moderate and livable temperature conditions in Western 
Europe~\cite{trenberth2019observation}. The AMOC transports warmer, upper waters in 
the Atlantic northward and returns colder, deeper waters 
southward~\cite{jackson2022evolution}. Studies suggested that, since about 30 years 
ago, there has been a tendency for the AMOC to 
weaken~\cite{biastoch2008causes,yeager2014origins}. At the present, the AMOC is still 
in a ``healthy'' steady state that maintains a stable circulation of the pertinent 
ocean flows. A potential halt of the circulation would signify a collapse of the AMOC
with dire consequences, which corresponds to another stable steady state of the 
underlying dynamical system. Such a collapse meets the criterion of a tipping point,
i.e., a transition from one stable steady state to another. 

\begin{figure*} [ht!]
\centering
\includegraphics[width=\linewidth]{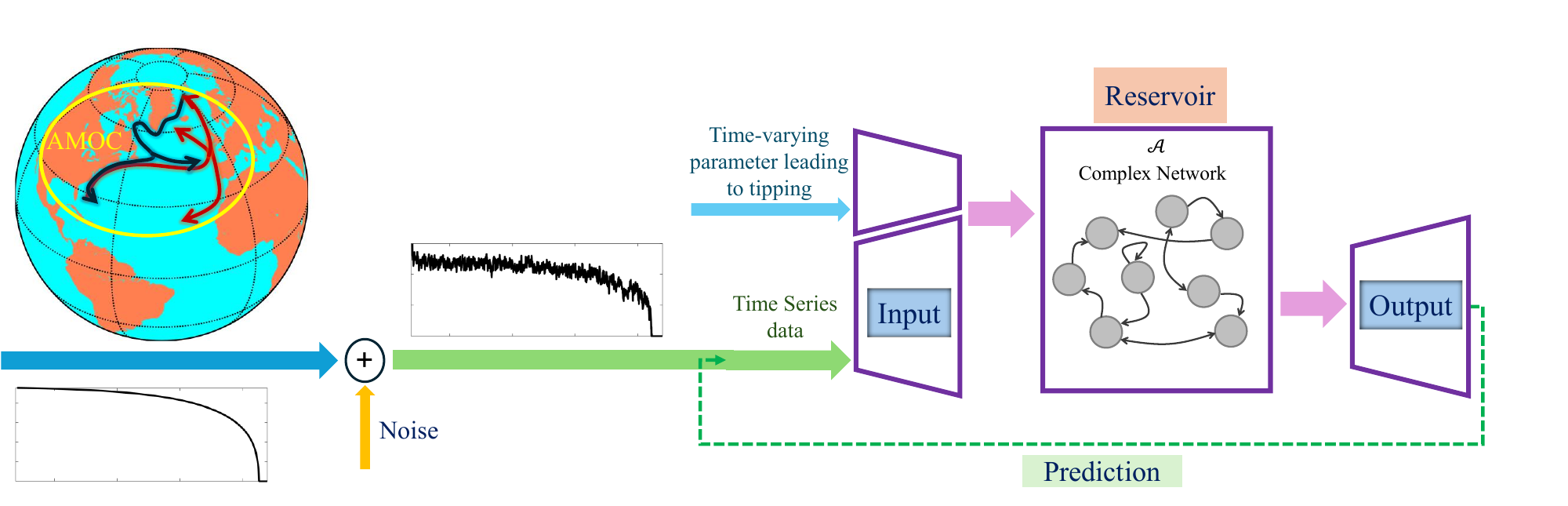}
\caption{Schematic illustration of the machine-learning framework for anticipating 
tipping in nonautonomous dynamical systems. The system begins in a stable steady 
state with no deterministic oscillations in the dynamical variables. Dynamic noise 
is leveraged to perturb the system, enabling the machine-learning model to detect 
changes and predict the tipping point even when the system is in a parameter regime 
prior to tipping.}
\label{fig:schematic}
\end{figure*}

It is worth emphasizing that the phenomenon of tipping in its original 
context~\cite{Scheffer:2004,Schefferetal:2009,Scheffer:2010,WH:2010,DG:2010,CLLLA:2012,BH:2012,DVKG:2012} is characteristically distinct from the more commonly studied 
critical transitions from an oscillatory state to some final state. Examples of 
such transitions include a crisis through which a chaotic attractor is destroyed 
and replaced by a chaotic transient~\cite{GOY:1983}, the onset of synchronization 
from a desynchronization state~\cite{FKLW:2021}, amplitude death~\cite{XKSL:2021}, 
and the encountering with a periodical window~\cite{PCGPO:2021}. While machine 
learning, in particular reservoir computing, has been applied to predicting such 
critical transitions~\cite{KFGL:2021a,KFGL:2021b,PO:2022,KWGHL:2023}, a shared 
characteristic among the existing works is the system's oscillatory behavior before 
the transition. This is advantageous because the time series for training the neural 
networks contain the temporal variations necessary for the machine to learn the 
dynamics of the system. Predicting a tipping point is significantly more challenging 
because, prior to the tipping, the system is in a stable steady state with no 
oscillations in the dynamical variables.

Our solution for machine-learning based prediction of tipping is exploiting dynamic 
noise. In particular, time series measured from real-world systems are noisy, and 
the inherent random oscillations are naturally suited for machine-learning training. 
In developing a machine-learning prediction framework, synthetic data are needed for 
validation. In this case, we generate time series with random perturbations about 
the deterministic steady state through stochastic dynamical modeling. While the 
presence of noise may potentially compromise the prediction accuracy, it serves a 
dual purpose by facilitating an adequate exploration of the phase space by the neural 
network dynamics, unveiling latent features that would otherwise remain obscured 
under noise-free conditions. A recent work has established that dynamical noise 
and/or measurement noise in the training dataset can be beneficial to the training 
process through a stochastic-resonance mechanism~\cite{ZKL:2023}. In addition, 
optimal calibration of noise levels can mitigate the risk of overfitting and promote 
generalization, allowing the reservoir computer to adapt to varying environmental 
conditions and data distributions. Incorporating a parameter channel into reservoir 
computing~\cite{KFGL:2021a} to accommodate the time-varying parameter, we demonstrate 
that the reservoir computer can be trained to predict the occurrence of tipping in 
the future. To show the efficacy of our prediction framework, we present examples 
from climatic systems and ecological networks. For the problem of anticipating a 
potential collapse of the AMOC with synthetic and currently available empirical 
data, our machine-learning scheme places a collapse window spanning from 2040 to 
2065, in consistence with the results in the current literature.

\section*{Results}

We present results on anticipating a potential collapse of the AMOC here in the 
main text, while leaving other case studies to Supplementary Information (SI).
Our machine-learning framework is designed to tackle the challenge of anticipating 
tipping in nonautonomous dynamical systems in general, as shown in 
Fig.~\ref{fig:schematic} where, prior to tipping, the system is in a stable steady 
state with no deterministic oscillations in the dynamical variables. We use four 
types of data: (1) synthetic data from one-dimensional (1D) AMOC fingerprint model, 
(2) synthetic data from a 2D conceptual model of AMOC, (3) synthetic data from the 
Community Earth System Model, and (4) empirical AMOC fingerprint data. As will be 
demonstrated, the predicted time window of a potential future AMOC collapse from 
the four types of data are consistent with each other.

\begin{figure} [ht!]
\centering
\includegraphics[width=\linewidth]{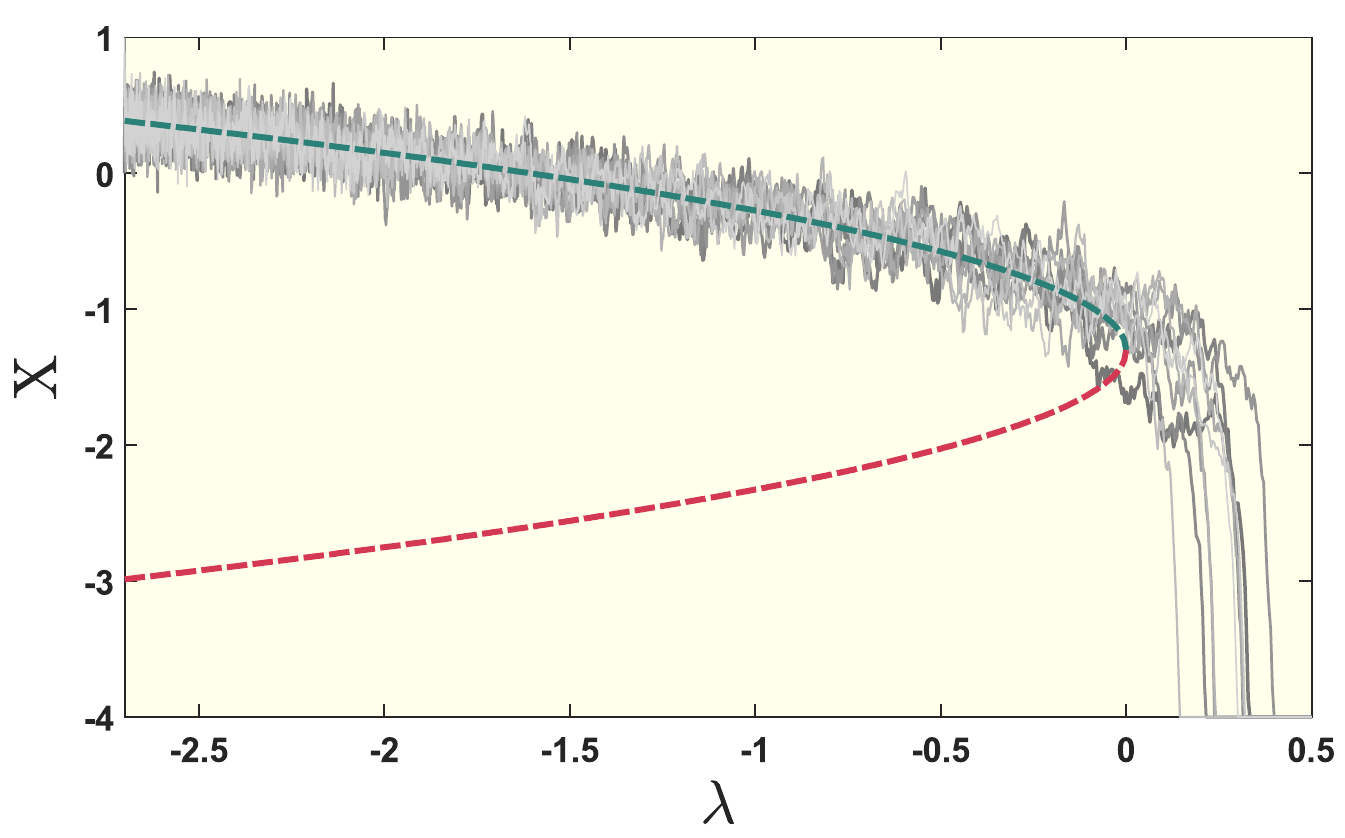}
\caption{Random realizations of a tipping point transition in the 1D stochastic AMOC 
fingerprint model~\eqref{Eq: 1D AMOC}. The bifurcation parameter $\lambda(t)$ 
increases exponentially with time, while other parameters are the best-estimated 
values extracted from the empirical fingerprint data~\cite{ditlevsen2023warning}: 
$A=0.95$, $m=-1.3$, $\lambda_0 = -2.7$, $\sigma = 0.3$, $t_0 = 1924$, and 
$\lambda_c =0$. Ten random realizations are shown, with the dashed green and red 
curves indicating the stable and unstable equilibria. In the underlying deterministic 
system, a backward saddle-node bifurcation and hence a tipping point occurs at 
$\lambda_c = 0$. In the presence of stochastic driving, the value of $\lambda$ at 
which the system collapses, characterized by the dynamical variable $X$'s approaching 
a large negative value, varies among the realizations, but they are near $\lambda=0$ 
on the positive side.}
\label{fig:1D_fingerprint}
\end{figure}

\begin{figure*} [ht!]
\centering
\includegraphics[width=\linewidth]{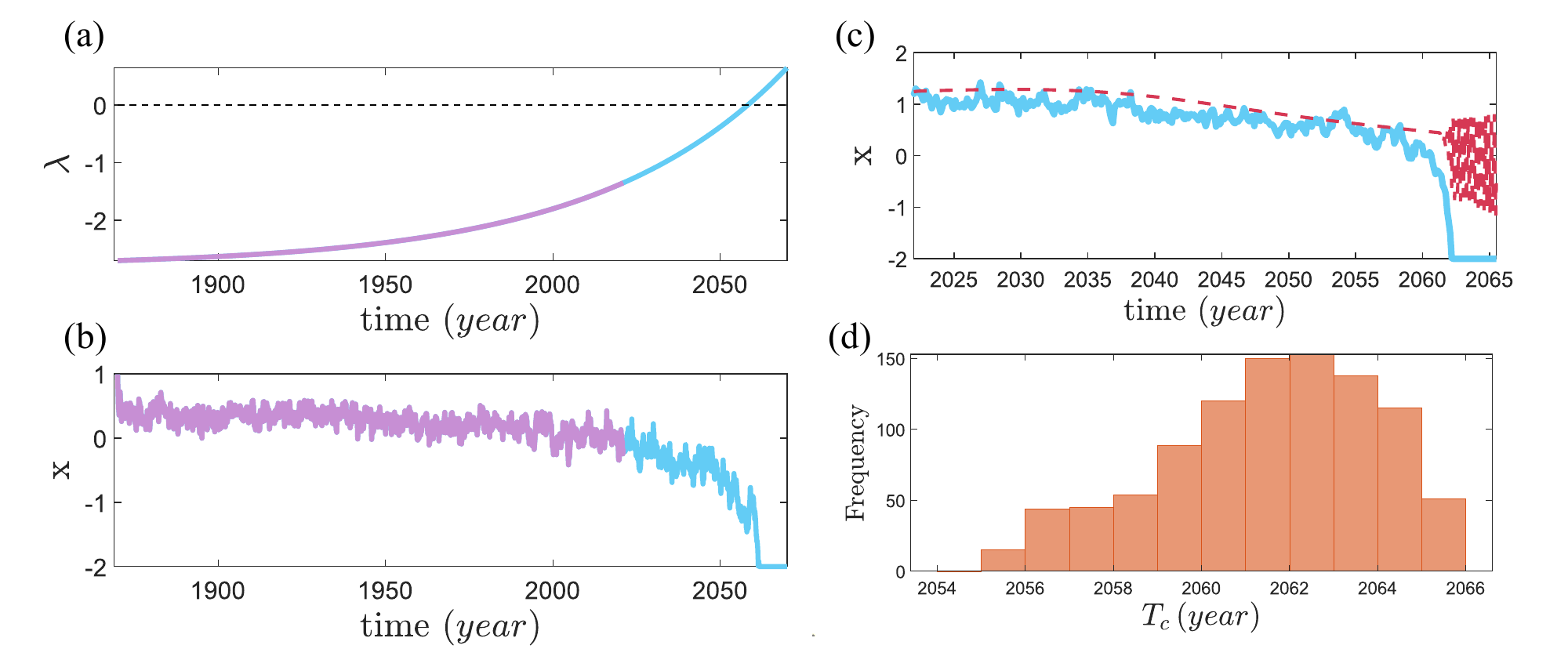}
\caption{Reservoir-computing prediction of the time window of AMOC collapse from 
the 1D time-dependent fingerprint model. (a) The exponential growth with time of the 
bifurcation parameter $\lambda (t)$, which starts from the value $\lambda_0 = -2.7$ 
in the year 1870. The horizontal dashed line indicates the tipping point 
$\lambda_c=0$. (b) A realization of the time series $X(t)$, where the purple 
(blue) segment represents the training and testing (validation and prediction) data, 
respectively. (c) The testing data (blue) and reservoir-computing prediction (red) 
in the time window from year 2022 to year 2065. For this particular realization, 
the AMOC variable $X(t)$ collapses between the years 2062 and 2063 (blue, real data). 
The reservoir computer predicts an abnormal behavior in $X(t)$ at about the same 
critical time $T_c$, signifying a tipping point. (d) Histogram of the predicted AMOC 
collapse time $T_c$ obtained from 1000 machine realizations. Tipping is likely to 
occur between year 2055 and year 2066.}
\label{fig:RC_FP_Prediction}
\end{figure*}

\paragraph*{\bf Anticipating AMOC collapse from 1D synthetic fingerprint data.}
Due to the difficulty of continuously monitoring the AMOC and the limited availability 
of long-term observational data, analyzing certain fingerprints of the AMOC provides 
a viable method to gain insights~\cite{Jackson2020}. For example, sea surface
temperature (SST) has been employed as a promising proxy for assessing the AMOC
strength~\cite{Robson2013,Zhang2017,Haskins2019,BenYami2023}. Quite recently, a
1D stochastic SST model~\cite{ditlevsen2023warning} with parameter values estimated 
from the real data was constructed to understand the tipping dynamics of the AMOC.
It was suggested that the AMOC may be approaching a potential collapse through a 
tipping point, which can occur as early as 2025. The model is described by the 
following stochastic nonlinear differential equation with a generic bifurcation 
parameter $\lambda$:
\begin{equation}\label{Eq: 1D AMOC}
    \dot{X}_t = -[A(X_t - m)^2 + \lambda)] + \sigma dB_t,
\end{equation}
where $X_t$ is a stochastic dynamical variable exhibiting a tipping transition, 
$A$ is a time scale parameter, $m$ is defined as $\mu - \sqrt{|\lambda|/A}$ with 
$\mu$ representing the stable fixed point of the process, $B_t$ is a Brownian 
motion, and $\sigma$ is the noise amplitude. Initially, the system is in a 
statistically stable state with constant $\lambda = \lambda_0$. At time $t_0$, 
$\lambda$ begins to increase toward the critical point $\lambda_c$. As $\lambda$ 
increases, the dynamical variable $X_t$ exhibits fluctuations but its mean value 
decreases continuously. Despite the fluctuations, $X_t$ eventually collapses to a 
large negative value, signifying the collapse of the AMOC. For fixed model 
parameters at the most likely estimated values from the AMOC fingerprint data 
($A=0.95$, $m=-1.3$, $\lambda_0 = -2.7$, and $\sigma = 0.3$, and 
$t_0 = {\rm Year} \ 1924$), the underlying deterministic system exhibits a backward 
saddle-node bifurcation, corresponding to the coalescing point of the stable and 
unstable equilibrium points, as shown in Fig.~\ref{fig:1D_fingerprint}, where a 
tipping point occurs at $\lambda_c = 0$ (see Sec.~III in SI for details). 

Climate change is a driving force to slow down and eventually halt the AMOC, making
the underlying dynamical system nonautonomous. The nonautonomous version of 
Eq.~(\ref{Eq: 1D AMOC}) can be obtained by making the bifurcation parameter 
$\lambda$ time-dependent. In particular, the impact of climate change was 
modeled~\cite{ditlevsen2023warning} by an exponential increase in $\lambda (t)$ 
with time from some initial value $\lambda_0 < 0$, as illustrated in 
Fig.~\ref{fig:RC_FP_Prediction}(a). As $\lambda$ increases towards the bifurcation 
point $\lambda_c = 0$, the system approaches a tipping point at the time $T_c$. In 
the time interval, $[0,T_c]$, the AMOC variable $X(t)$ fluctuates about the stable 
equilibrium. After the tipping at $T_c$, $X(t)$ rapidly decreases to a large negative 
value, signifying the AMOC collapse. The value of the tipping time $T_c$ varies 
across different realizations. 

We now demonstrate that a trained reservoir computer is able to predict the tipping 
time $T_c$. For each realization, we divide the data into two distinct segments: 
training and testing, as highlighted in Figs.~\ref{fig:RC_FP_Prediction}(a) and 
\ref{fig:RC_FP_Prediction}(b) in purple and blue, respectively, where the end of the 
purple data segment marks the present time (in year). Note that, up to the present 
time, the AMOC has been stable, where the dynamical variable $X(t)$ fluctuates about 
the healthy stable steady state. If there was no noise, $X(t)$ would be a smooth and 
a slowly decreasing function of time, as exemplified in Fig.~\ref{fig:1D_fingerprint},
and it is not possible to train the reservoir computer with the non-oscillatory time 
series. What makes training possible is noise rendering oscillatory and random the 
time series $X(t)$. Figure~\ref{fig:RC_FP_Prediction}(c) presents one prediction run, 
where the blue trace is the testing data in the time interval between now and year
2065 (the ground truth), and the dashed red trace is the reservoir-computing 
prediction. For this particular realization, the predicted AMOC collapse time is 
between the years 2062 and 2063. At about the same time, the predicted $X(t)$ exhibits 
an abnormal behavior that is drastically and characteristically different from that 
prior to the tipping, indicating a successful prediction of the tipping point. Note 
that, since the reservoir computer has never ``seen'' the blue testing data segment 
that includes the collapse of $X(t)$ to some negative value, it is not possible for 
the machine to predict the value of $X(t)$ after tipping. Nevertheless, the predicted 
abnormal behavior is indicative of some critical behavior in the system. 
Figure~\ref{fig:RC_FP_Prediction}(d) shows a histogram of the predicted values of 
$T_c$ from 1000 reservoir-computing realizations. For the 1D AMOC fingerprint model, 
the parameter-adaptable reservoir computer (detailed in {\bf Methods}) predicts 
that a collapse of the AMOC is likely to occur between the years 2055 and 2066, which 
is consistent with the result in Ref.~\cite{ditlevsen2023warning}.

The histogram of the collapse time $T_c$ in Fig.~\ref{fig:RC_FP_Prediction}(d) was 
obtained from 1000 machine realizations, but the training and testing data are from 
one specific realization of the 1D AMOC fingerprint model. For different model 
realizations, the tipping time $T_c$ is different, so are the predictions. 
Table~\ref{tab:T_c_statistics} lists the prediction results from 10 model 
realizations. It can be seen that in all cases, the predicted mean value of the 
collapse year is close to that of the original data, providing further validation 
of our reservoir-computing prediction scheme. 

\begin{table} [ht!]
\centering
\caption{Predicted tipping time from 20 synthetic datasets}
\label{tab:T_c_statistics}
\begin{tabular}{|c|c|c|c|}
\hline
\multirow{2}{*}{Dataset} & \multirow{2}{*}{Model $T_c$} & \multicolumn{2}{c|}{$T_c$ from 1000 machine realizations}\\
    \cline{3-4}
	& & \quad \quad Mean\quad \quad \quad & Std (years) \\ \hline
     1 & 2061 & 2062 & 4\\
     2 & 2054 & 2057 & 6\\
     3 & 2056 & 2057 & 3\\
     4 & 2070 & 2069 & 6\\
     5 & 2064 & 2062 & 5\\
     6 & 2059 & 2060 & 4\\
     7 & 2058 & 2060 & 5\\
     8 & 2062 & 2061 & 4\\
     9 & 2060 & 2062 & 6\\
     10 & 2065 & 2066 & 4\\
     \hline
\end{tabular}
\end{table}

\paragraph*{\bf Predicting AMOC collapse based on synthetic data from a 2D conceptual 
model.} A recent study~\cite{lohmann2021risk} addressed the phenomenon of tipping 
within climate systems, providing insights into how time-varying parameters can lead 
to abrupt and potentially catastrophic transitions in the climate. The mechanism of 
tipping was illustrated using a 2D conceptual model~\cite{lohmann2021risk} with two 
state variables, denoted as $x$ and $y$. The dynamics produced by this model closely 
resemble those observed in the ocean model. The 2D model is described by the following 
equations: 
\begin{align} \label{eq:AMOC_2D}
    \frac{dx}{dt} &= (-r^4 + 2r^2 - \beta)x - \omega \hat{y},\\
    \frac{d\hat{y}}{dt} &=(-r^4 + 2r^2 - \beta)\hat{y} + \omega x,
\end{align}
where $r^2 = x^2 + \hat{y}^2$, and $\hat{y}$ is defined as $y - \gamma \beta$. The 
parameter $\beta$ is a bifurcation parameter describing the freshwater forcing 
parameter, while $\omega$ represents the frequency. The dependence of $y$ on 
$\beta$ is parameterized by $\gamma$. The system possesses one stable and one 
unstable limit cycle for $\beta<1$. For $\beta = 1$, the limit cycles merge and 
disappear in a saddle-node bifurcation~\cite{lohmann2021risk}.

\begin{figure*} [ht!]
\centering
\includegraphics[width=\linewidth]{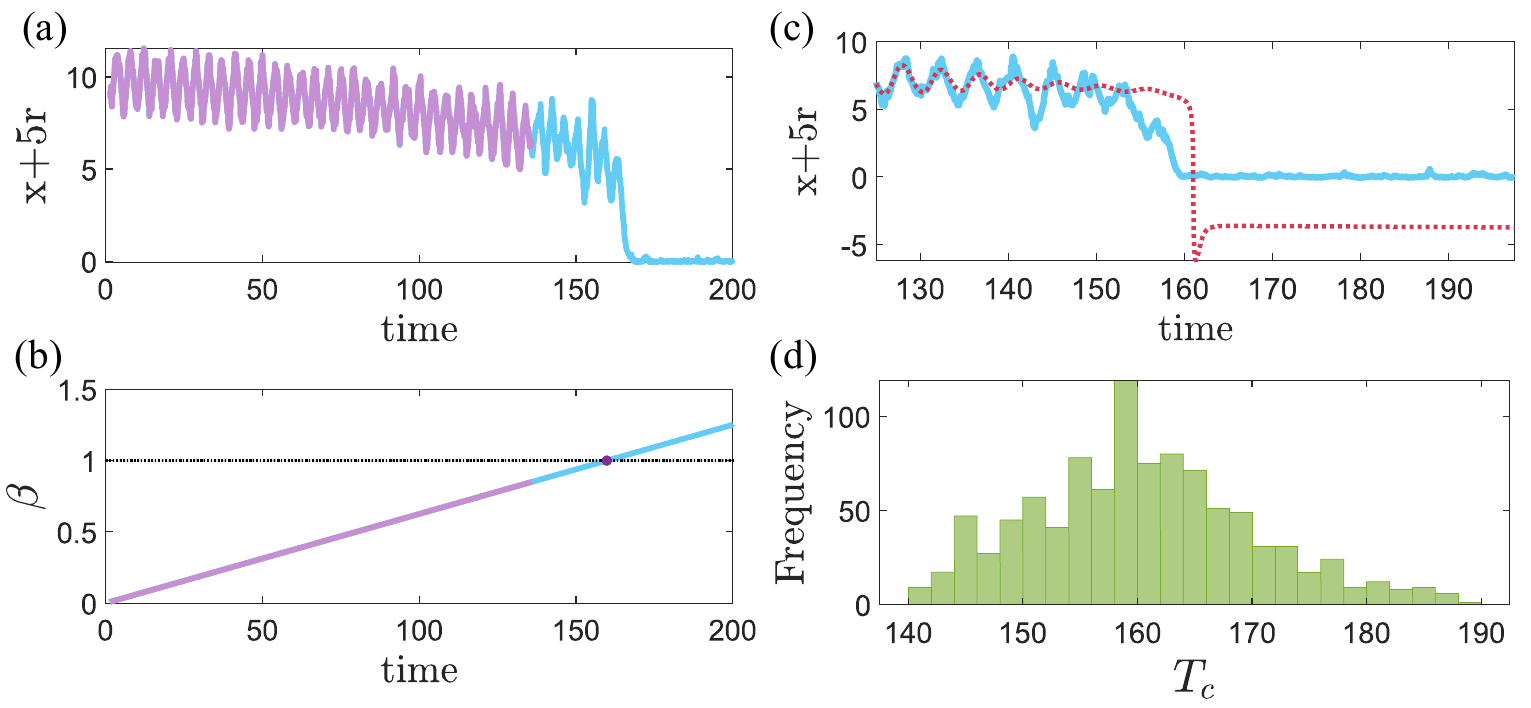}
\caption{Reservoir-computing prediction of the time window of AMOC collapse from the 
2D time-dependent conceptual AMOC model. (a) One realization of 2D conceptual AMOC
model for $\gamma = 3$ and $\sigma = 0.1$. (b) Time-varying freshwater forcing 
parameter $\beta$. (c) An example of testing data (solid blue trace) and 
reservoir-computing prediction (dash-dotted red trace). (d) Histogram of the predicted 
critical point from $1000$ random reservoir realizations.}
\label{fig:2d_conceptual}
\end{figure*}

To make the dynamical system (\ref{eq:AMOC_2D}) nonautonomous, we assume that the
bifurcation parameter $\beta$ is time-dependent: starting from an initial value 
$\beta_0$, it linearly increases towards a critical value $\beta_c$. Dynamical noise
is introduced into the system by additive stochastic terms in (\ref{eq:AMOC_2D}) 
of independent Gaussian random processes of zero mean and amplitude $\sigma$. 
Figures~\ref{fig:2d_conceptual}(a) and \ref{fig:2d_conceptual}(b) show a single 
realization of the evolution of the conceptual AMOC fingerprint ($x+5r$) and the 
$\beta$ parameter over time, respectively. As $\beta$ increases, a tipping point  
occurs at $\lambda_c = 0$, where the system undergoes a sudden shift in the attractor, 
transitioning from random oscillations about the stable limit cycle to a stable 
equilibrium. The specific tipping time varies among the independent stochastic
realizations.

\begin{figure*} [ht!]
\centering
\includegraphics[width=\linewidth]{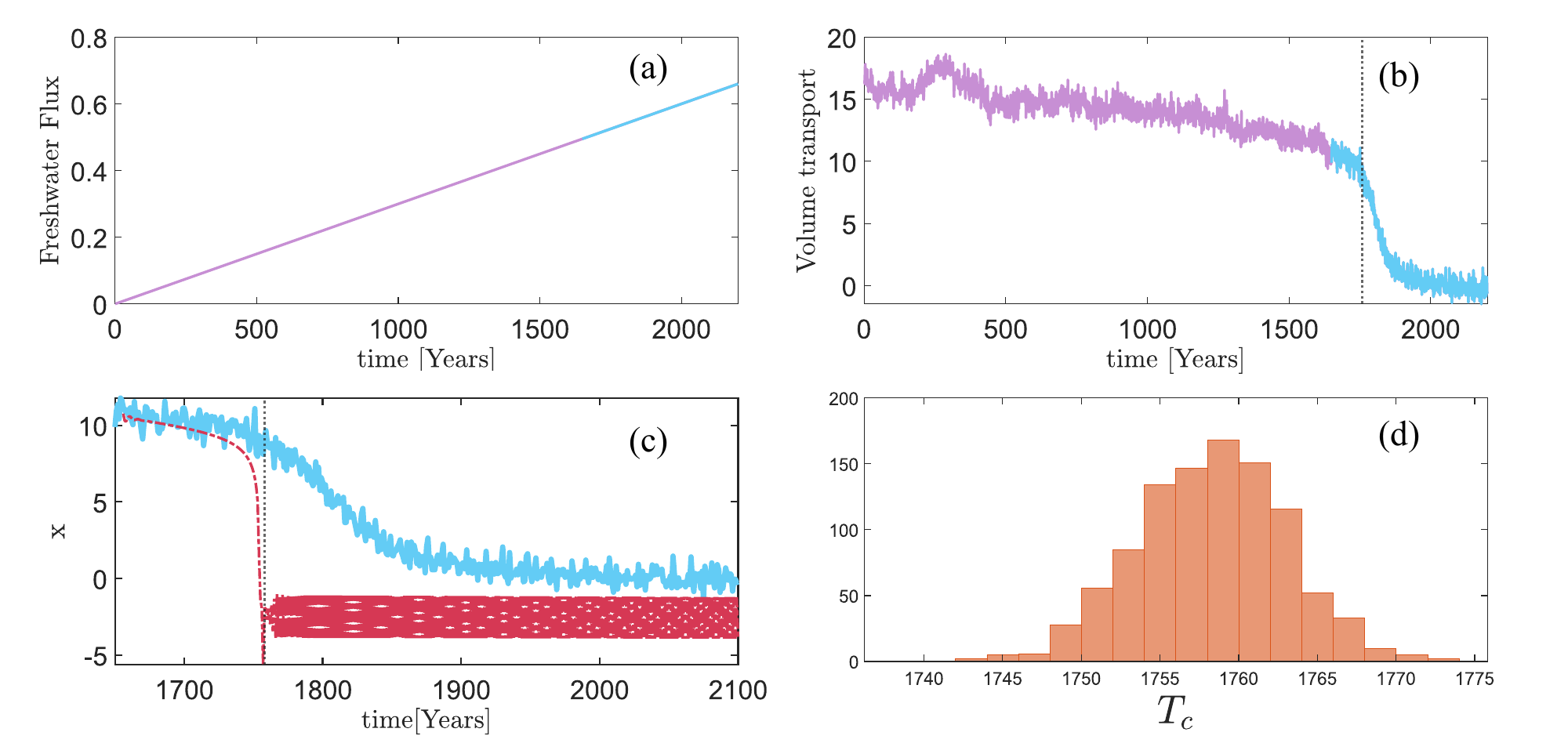}
\caption{Reservoir-computing prediction of the time window of potential AMOC collapse 
from the CESM synthetic data. (a) Linear time-varying freshwater flux. (b) AMOC 
strength, where the purple and blue segments represent the training and testing 
(validation and prediction) data, respectively. The horizontal dashed line indicates 
the tipping point $T_c = 1758$. (c) Testing data (blue) and reservoir-computing 
prediction (red). AMOC strength ($x(t)$) collapses at model year $1758$ (blue, real 
data). The reservoir computer predicts an abnormal behavior in $x(t)$ at about the 
same critical time $T_c$, signifying a tipping point. (d) Histogram of the predicted 
AMOC collapse time $T_c$ obtained from 1000 reservoir network realizations. Tipping 
is likely to occur between model years 1740 and 1775.}
\label{Revised_1}
\end{figure*}

To anticipate a possible tipping point, we partition the data into two sets: training 
data (highlighted in purple) and testing data (in blue). The training data consists 
of a portion of the time series of $x+5r$ as input and $\beta$ as a control parameter 
associated with oscillations about the stable limit cycle. During the testing phase, 
the trained reservoir computer is employed alongside the remaining control parameter 
data to predict the tipping. Figure~\ref{fig:2d_conceptual}(c) shows an illustrative
example, where the real testing data are in blue and the corresponding predicted data 
are represented by red. Training data are collected for parameter values 
$\beta \in [0.01, 0.79]$. To ensure the prediction efficacy, we repeat the whole 
process for $1000$ random realizations of the reservoir computer. 
Figure~\ref{fig:2d_conceptual}(d) presents a histogram of the anticipated tipping 
point values. In all the realizations, a tipping point is anticipated to occur in 
the future within the time interval $T_c \in [140,190]$, which contains the ground
truth value $T_c = 159$ from direct simulation of the 2D stochastic system.

\paragraph*{\bf Predicting AMOC collapse using the synthetic data from the Community 
Earth System Model.} In a quite recent study of the Community Earth System Model 
(CESM)~\cite{vanWesten2024}, an AMOC tipping event with significant climate 
consequences was revealed. CESM is a coupled climate model for simulating various 
components of the Earth's climate system simultaneously, making it possible to 
explore the dynamics under the past, present, and future climate conditions. An 
analysis of the output data of CESM revealed a tipping point as characterized by the 
minimum of the AMOC-induced freshwater transport at the Southern boundary of the 
Atlantic~\cite{vanWesten2024}.

We use the simulated AMOC strength data from Ref.~\cite{vanWesten2024} to test our 
reservoir-computing based framework for predicting tipping. In the CESM model, the 
freshwater flux forcing ($F_H$) linearly increases at the rate $3 \times 10^{-4}$ 
Sv year-1 until the model year $2200$, where a maximum of $F_H = 0.66$ Sv is reached, 
as shown in Fig.~\ref{Revised_1}(a). The AMOC strength, defined as the total 
Meridional volume transport, is shown in Fig.~\ref{Revised_1}(b), where the 
vertical dashed line indicates a tipping point $T_c = 1758$. The purple segment is 
used to train the reservoir computer and the blue segment is the testing data. The 
reservoir-computing output is shown in red in Fig.~\ref{Revised_1}(c), where an 
abnormal behavior in $x(t)$ occurs at about the same critical time $T_c$ as the 
model tipping time. To characterize the prediction performance, we repeat the 
process using $1000$ machine-learning realizations. The resulting histogram of the 
predicted AMOC collapse time $T_c$ is shown in Fig.~\ref{Revised_1}(d), which 
indicates that tipping is likely to occur between the model years $1740$ and $1775$.

\begin{figure} [ht!]
\centering
\includegraphics[width=\linewidth]{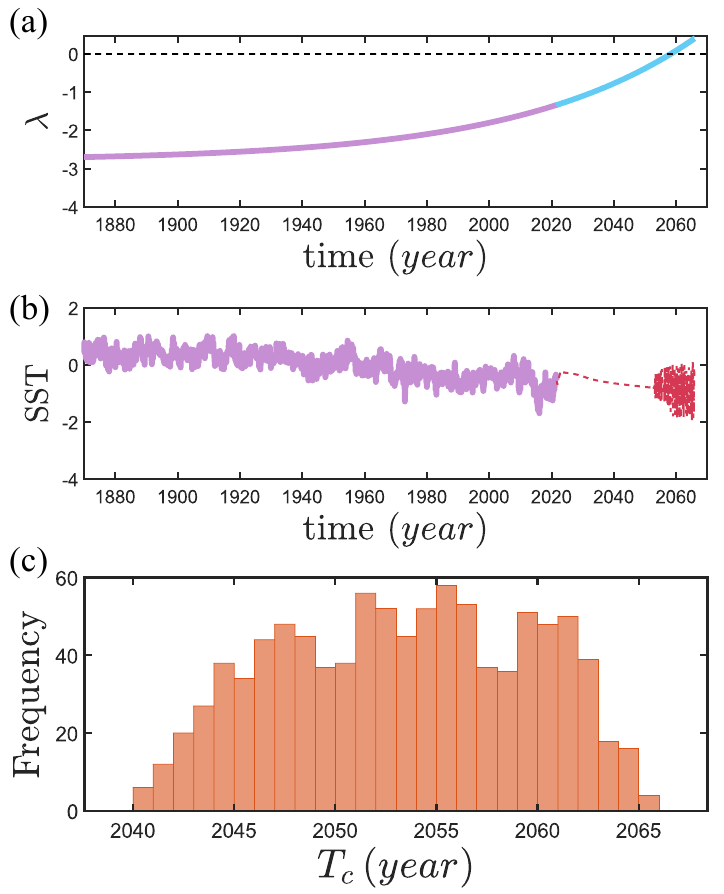}
\caption{Reservoir-computing prediction of AMOC collapse time using empirical 
fingerprint data. (a) An exponential growth of the bifurcation parameter $\lambda$ 
(from Ref.~\cite{ditlevsen2023warning}). (b) Available AMOC fingerprint SST data 
from year 1875 to the present year (purple). This data segment is noisy and employed 
in training the reservoir computer. The red trace is one example of the predicted 
SST behavior, which is smooth until the critical value $\lambda_c = 0$ for collapse 
is reached. (c) A histogram of the predicted AMOC collapse time from 1000 
reservoir-computing realizations. The time range of potential AMOC collapse is 
between the year 2040 and the year 2066.} 
\label{fig:Real_Data}
\end{figure}

\paragraph*{\bf Predicting AMOC collapse using empirical fingerprint data.} It is 
necessary to conduct tests using empirical AMOC data. We use AMOC fingerprint 
sea-surface temperature (SST) datasets with the same exponential growth of the 
bifurcation parameter~\cite{ditlevsen2023warning}, as shown in 
Fig.~\ref{fig:Real_Data}(a). Figure~\ref{fig:Real_Data}(b) shows a segment of the 
SST data up to the present time (in purple color), which is used for training, and 
a typical realization of the reservoir-computing predicted time series (red). Prior 
to reaching the critical point $\lambda_c = 0$, the predicted AMOC fingerprint 
exhibits a smooth behavior that is essentially a continuation of the training data, 
indicating no collapse. About $\lambda_c = 0$, the machine-learning prediction 
becomes highly irregular, signifying a collapse. Figure~\ref{fig:Real_Data}(c) shows 
a histogram of the predicted critical time of AMOC collapse from 1000 
reservoir-computing realizations. The range of possible collapse time is from year 
2040 to year 2066, with the median around year 2053. This result is consistent with 
those in Fig.~\ref{fig:RC_FP_Prediction}(d) and in Ref.~\cite{ditlevsen2023warning}.

To further demonstrate the generality and power of our parameter-adaptable reservoir 
computing framework for predicting tipping in complex and nonautonomous dynamical 
systems, we tested the following datasets from alternative AMOC models and ecological 
networks that exhibit a tipping point in the conventional sense of coexisting stable 
fixed-point attractors: (1) a three-box AMOC model, (2) mutualistic pollinator-plant 
networks, (3) a plant-herbivore model, and (4) a climate model. For models (1-3), a 
bifurcation parameter is assumed to vary continuously with time but for model (4), 
the observed time series are collected from a sequence of different constants or 
nearly constant parameter values. The details of the models and the prediction 
results are presented in SI.

\section*{Discussion}

Recent years have witnessed significant efforts in developing machine-learning
models for predicting critical transitions in nonlinear dynamical systems. A tacit 
assumption in these works is that oscillatory time series are available for training
the neural network. For critical transitions such as crises, synchronization onset
and amplitude death, this requirement can indeed be met. Tipping, by its historical 
origin from nonlinear ecosystems, is a different type of critical transition in that 
the system is in some sort of stable steady state before and after the transition. 
In a deterministic system, the available time series are non-oscillatory. The lack of 
pre-tipping oscillations means that the usual temporal variations used for training 
are absent, making it significantly more challenging to predict the impending shift 
using machine learning (see {\bf Methods} for more details). This is a reason that 
most previous works used {\em detection-based} approaches to extracting early warning 
signals or features from observed time series before the tipping.

We developed a {\em forecasting-based}, machine-learning framework to anticipate 
tipping in nonautonomous dynamical systems by taking advantage of noise. Our 
prediction is based on the presently available time-series data in a stable steady 
state but under the influence of noise. For synthetic data from a dynamical model,
we incorporate stochasticity into the governing equations to generate noisy time
series. For empirical data from the real word, most likely they are already noisy.
In any case, the random oscillations associated with the noisy time series make it
possible to train a machine-learning model, such as reservoir computing. We tested
our parameter adaptable reservoir-computing scheme on a variety of systems from 
diverse fields, all sharing the common tipping scenario: sudden transition from 
one steady state to another as a bifurcation parameter passes through a critical 
point. The main application is predicting the potential collapse of the AMOC.
Using simulated and empirical fingerprint data, our results suggest that the AMOC 
could halt in a time window centered about the year 2055, with the earliest possible 
occurrence in year 2040. These are consistent with the recent results based on a 
statistical optimization approach~\cite{ditlevsen2023warning}. 

Our machine-learning method is predicated on the availability of a known time-varying 
parameter that drives the system towards tipping~\cite{KFGL:2021a}. A simple 
assumption, in the absence of detailed information, is that the control parameter 
changes gradually and approaches its unknown critical value linearly over 
time~\cite{ditlevsen2023warning,vanWesten2024}. However, this linear assumption does 
not fully capture the complexities of real-world scenarios. The exact nature of the 
time-varying parameter and its real-time changes remain ambiguous. It has been 
found that the AMOC is sensitive to variations in the ocean's freshwater 
forcing~\cite{Taherkhani2020,ditlevsen2023warning,vanWesten2024} that can manifest 
through surface freshwater fluxes such as precipitation or through the input of 
freshwater from river runoff and ice melt, including significant contributions from 
the Greenland Ice Sheet. More sophisticated models suggested that the freshwater flux 
exhibits a quasi-exponential behavior~\cite{Swart2013,Gorte2023}. We have studied
the case where the time-varying parameter $\lambda(t)$ changes exponentially over 
time. This assumption aligns more closely with observed behaviors and enhances the 
accuracy of our machine-learning prediction, enabling better anticipation of a 
potential tipping of the AMOC.

Another technical issue is whether the predicted collapse is merely an artifact 
caused by the reservoir computer operating in an untrained parameter region. To 
address this concern, we conducted simulations to test the extrapolation results on 
the other side of the untrained parameter region using synthetic AMOC data. 
For example, we tested $\lambda$ values smaller than those in the trained parameter 
interval $\lambda \in (-2.5,-1.5)$ in the 1D AMOC model. In the simulations, we 
conducted 500 testing trials. For all these trials, no collapse was observed in 
this parameter region. The reservoir computer consistently and persistently predicted 
that the system remained in a healthy, stable steady state with no indication of an 
impending collapse. This consistent behavior across numerous trials strongly suggests 
that the predicted collapse is not an artifact of the reservoir computer functioning 
outside its trained parameter region, but rather robust prediction of the 
system's dynamics.

\section*{Methods}

\paragraph*{\bf Nonlinear dynamical mechanism of tipping.}
In nonlinear dynamics, a typical bifurcation leading to tipping is the forward or 
backward saddle-node bifurcation. Consider the situation of two coexisting stable 
steady states (or attractors): a normal ``healthy'' state and a catastrophic or 
``low'' state, where each attractor has its own basin of attraction. As the 
bifurcation parameter increases with time, the healthy attractor can disappear 
through a backward saddle-node bifurcation, after which the low state is the only 
attractor in the phase space, signifying a tipping point. In the past, considerable 
efforts were devoted to anticipating tipping by identifying early warning indicators 
or signals~\cite{SBBBCDHNRS:2009,BRH:2013,LeemputEtal:2014,Boers:2018,BSPSLAB:2021}. 
A known phenomenon is enhanced fluctuations where, as the tipping point is approached, 
the variances of the measured values of the dynamical variables tend to increase. The 
reason is that, as the system moves toward a fold bifurcation, the dominant eigenvalue 
of the Jacobian matrix evaluated at the steady state approaches zero, making the 
landscape flatter and closer to a random walk about the steady-state attractor. Small 
noise will then generate large deviations of the trajectory from the attractor. In a 
recent work, a deep-learning based time-series classification scheme was introduced 
to determine if a tipping event is about to occur and the 
bifurcation~\cite{BSPSLAB:2021}.

\paragraph*{\bf Challenges with anticipating tipping.}
Oscillatory behaviors in the data in the pre-critical regime have the benefit of 
system trajectory's visiting a substantial portion of the phase space, thereby 
facilitating training by enabling the neural network to effectively learn the 
phase-space behavior or the dynamical climate of the target system. Differing from 
existing works on predicting critical transitions from an oscillatory dynamical state 
to a collapsed state, we aim to predict tipping from one stable steady state to 
another. In a noise-free situation, in the pre-tipping regime the system is in a 
stable steady state without oscillations in its dynamical variables. Introducing
stochasticity or noise into the system leading to randomly oscillating dynamical 
variables provides a solution for neural-network training. We exploited dynamic noise 
in the data for training, where validation and hyperparameter optimization are 
performed based on data in the pre-critical regime. During the test or prediction 
phase, the reservoir computer operates as a closed-loop, deterministic dynamical 
system capable of predicting how the dynamical climate of the system changes with 
the time-varying bifurcation parameter. Since no data from the target system in the 
post-tipping regime were used for training (in a realistic situation, such data are 
not available), it is not possible for the reservoir computer to correctly predict 
the detailed system behavior after the tipping. However, the neural machine is 
capable of generating characteristic changes in the output variables at the tipping 
transition, making its anticipation possible.

\begin{figure} [ht!]
\centering
\includegraphics[width=\linewidth]{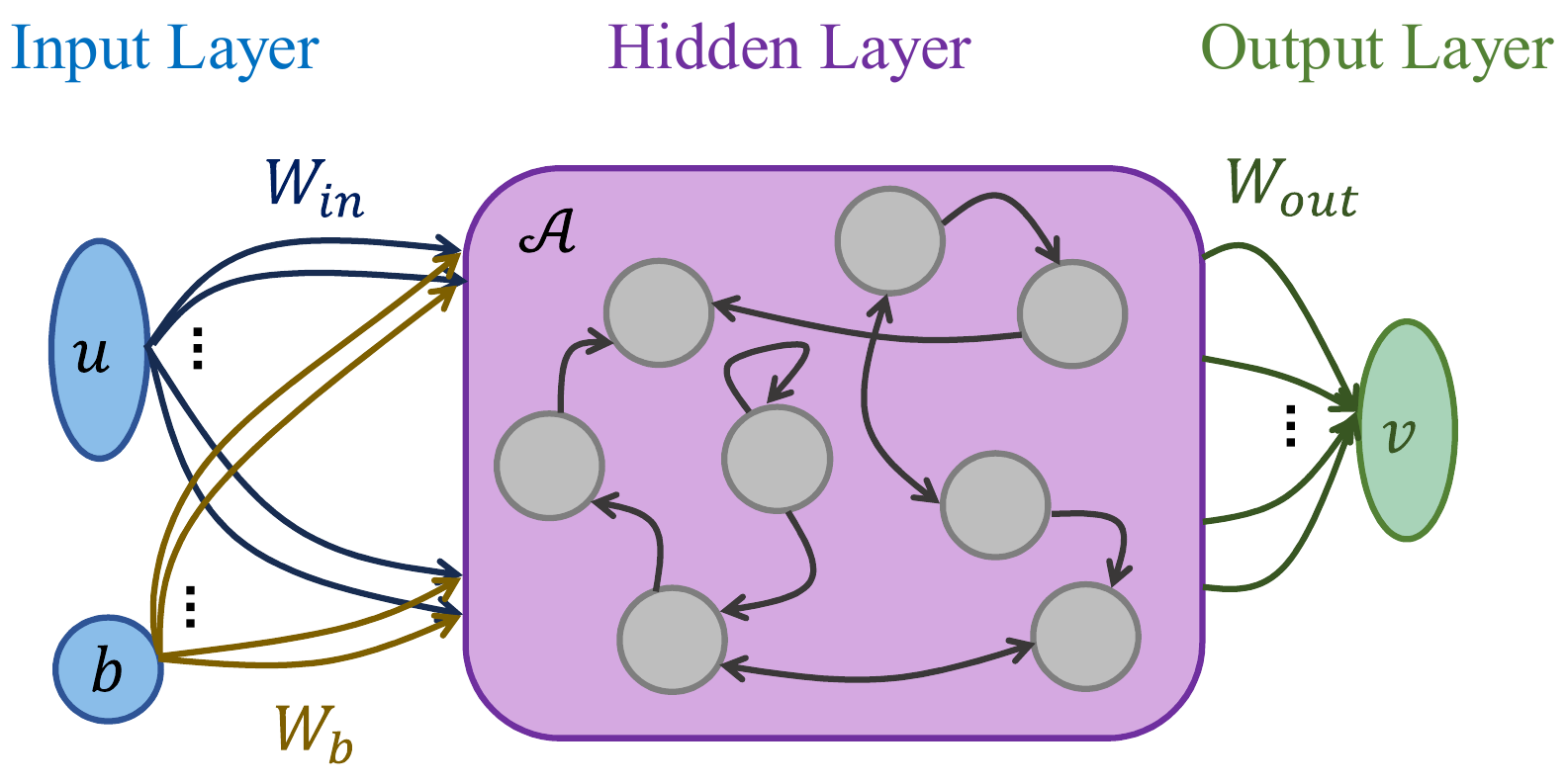}
\caption{Illustration of parameter-adaptable reservoir computing.}
\label{fig:RC}
\end{figure}

\paragraph*{\bf Parameter-adaptable reservoir computing.} We adopt 
parameter-adaptable reservoir computing~\cite{KFGL:2021a} for anticipating tipping. 
A basic reservoir computer comprises three layers: an input layer, a hidden recurrent 
layer, and an output layer. Figure~\ref{fig:RC} illustrates the basic structure of 
parameter-adaptable reservoir computing that extends conventional reservoir
computing by incorporating an additional parameter channel for the bifurcation 
parameter $b$. During the training, the input time series vector $\mathbf{u}(t)$ 
and the parameter $b$ are concurrently projected onto the hidden layer through the
time-series input matrix $W_{\text{in}}$ and the parameter input matrix $W_b$, 
respectively. The hidden layer consists of $N$ 1D dynamical neurons. Concatenating
the dynamical states of all the neurons leads to an $N$-dimensional vector - the 
hidden state $\mathbf{r}(t)$ at each time step. The neural network in the hidden 
layer is recurrent with the connection matrix $W_{\mathbf{r}}$ and short-term memory. 
The output matrix $W_{\text{out}}$ projects the hidden state $\mathbf{r}(t)$ to the 
output layer, generating the output vector $\mathbf{v}(t)$. The iteration equations 
of the parameter-adaptable reservoir computer are 
\begin{widetext}
\begin{align} \label{eq:RC1}
\mathbf{r}(t) &= (1-\alpha_r)\mathbf{r}(t-\Delta t) 
	+ \alpha_r \tanh \left[ W_{\mathbf{r}} \mathbf{r}(t - \Delta t) + W_{\text{in}}\mathbf{u}(t) + W_b(k_b b + b_b)\right], \\ \label{eq:RC2}
	\mathbf{v}(t) &= W_{\text{out}} \mathbf{f}(\mathbf{r}(t)), 
\end{align}
\end{widetext}
where $\alpha_r\in(0,1]$ is the leakage parameter defining a temporal scale of the 
reservoir network, $\Delta t$ is the time step, $\tanh(\cdot)$ is the hyperbolic 
tangent function serving as the nonlinear activation function in the hidden layer, 
$k_b$ and $b_b$ are the gain and bias of the parameter $b$, respectively, and 
$\mathbf{f}(\cdot)$ is a nonlinear output function of the reservoir computer.

A feature of the reservoir network is the random generation and the subsequent 
fixation of the input matrices $W_{\text{in}}$ and $W_b$, along with the recurrent 
network matrix $W_{\mathbf{r}}$. These matrices remain fixed during training 
with only the output matrix $W_{\text{out}}$ undergoing optimization. This design 
choice eliminates the need for back propagation in time during the training, 
alleviating computational cost and mitigating potential difficulties such as 
vanishing and exploding gradients. Following the random generation of the three 
matrices, the training process begins by inputting the time series $\mathbf{u}(t)$ 
and the corresponding control parameter $b$ through the input layer. The dynamical
evolution of the neural network follows Eq.~\eqref{eq:RC1}. This process is also 
referred to as the ``listening phase'' or ``echoing phase'', as if the driving 
training signals are echoing in the hidden state. During the training, 
Eq.~\eqref{eq:RC2} is not invoked as the output matrix has not been trained yet. 
Multiple trials of the time series data from the target system, each associated
with a distinct parameter value, are presented as the training data. Upon completion 
of the echoing phase for a specific trial for a particular $b$ value, the 
parameter-adaptable reservoir computer is re-initialized for a new echoing phase 
for another training parameter value. The hidden state behaviors observed during the 
training are recorded, whose variations are implicitly linked to the corresponding 
parameter value $b$ since it affects the dynamical evolution of the state in the
hidden layer.

Let the length of each trial of the training time series be $T_{\text{train}}$ 
(in the unit of the number of steps) and the number of trials of training with 
different parameter values be $n_b$ ($n_b=4$ in our work). The ``echoing'' 
results $\mathbf{r}(t)$ are concatenated into a matrix $R$ of dimensions 
$N \times n_b T_{\text{train}}$. Applying the nonlinear function 
$\mathbf{f}(\cdot)$, we obtain the transformed matrix $R'= \mathbf{f}(R)$ that 
captures the echoing hidden state for subsequent linear regression. A training 
target is essential. We focus on reservoir networks whose output represents one-step 
prediction, where $\mathbf{v}(t)$ is equal to $\mathbf{u}(t+\Delta)$, making 
the training target the stacking of all training time series with a one-step 
difference from the input data, denoted as $V$. Finally, a ridge regression is 
conducted between $R'$ and $V$ to determine the output matrix:
\begin{align}
    W_{\text{out}} = V\cdot R'^T ( R'\cdot R'^T +\beta_r I )^{-1}, \label{eq:regression}
\end{align}
where $\beta_r$ is the coefficient of $L-2$ regularization.

An alternative training approach involves supplying a time series with a 
non-constant parameter value, a nonstationary time series with the corresponding 
time-varying parameter $b(t)$ as the training data~\cite{KFGL:2021a}. This 
configuration is more practical in various scenarios and is employed in our
work.

Having successfully trained a parameter-adaptable reservoir computer, we can now 
use it to make predictions for a specific parameter value $b$ of interest, which
is in a parameter regime different from that for training. The reservoir computer 
autonomously extrapolates the learned dynamics during training, generating 
predictions of the system dynamics at some unobserved parameter value. The 
iteration equation during prediction is given by
\begin{widetext}
\begin{align}
	\mathbf{r}(t) &= (1-\alpha_r)\mathbf{r}(t-\Delta t) + \alpha_r \tanh (W_\text{r}\cdot \mathbf{r}(t - \Delta t) + W_{\text{in}}\mathbf{v}(t-\Delta t) + W_b(k_b b + b_b)), \label{eq:RCpredict1} \\
	\mathbf{v}(t) &= W_{\text{out}} \mathbf{f}(\mathbf{r}(t)),
\end{align}
\end{widetext}
where $\mathbf{u}(t)$ in Eq.~\eqref{eq:RC1} is replaced by $\mathbf{v}(t-\Delta t)$. 
Given the recurrent structure of reservoir computing, it is necessary to properly 
initialize the hidden states in order to make any predictions. As one may observe
from Eq.~\eqref{eq:RC1}, a previous state $\mathbf{r}(t-\Delta t)$ is needed to 
calculate $\mathbf{r}(t)$. For short-term validation, we initialize the prediction 
by replicating the final hidden state obtained during the training as a one-step 
previous state. This allows for a direct comparison of the validation result with 
the actual time series, facilitating the calculation of an error metric, such as 
the root-mean-square error (RMSE), as the validation error. To conduct long-term 
testing on the ``climate,'' where a more diverse ensemble of predictions is required 
to capture the behaviors of the target system, we introduce additional randomness. 
Specifically, we utilize a randomly selected short segment from the actual time 
series data to ``warm up'' the reservoir hidden state. To prevent the reservoir 
computer from becoming stuck in a single attractor, especially in the presence of 
multistability in the dynamics, we introduce additive observational noise to the 
``warm-up'' data.

\section*{Data Availability}
\vspace*{-0.1in}
The data generated in this study, including both the training time series 
and the weights of the reservoir computers, can be found in the repository:
https://github.com/SPanahi/RC$_{-}$Tipping;
The CESM data are also available from: https://github.com/RenevanWesten/SA-AMOC-Collapse/tree/main/Data

\vspace*{0.1in}

\section*{Code Availability}
\vspace*{-0.1in}
The codes used in this paper can be found in the repository: 
https://github.com/SPanahi/RC$_{-}$Tipping

\bibliographystyle{naturemag}
%\bibliography{R_tipping}

\section*{Acknowledgments}
\vspace*{-0.1in}
This work was supported by the Air Force Office of Scientific Research under Grant 
No.~FA9550-21-1-0438 and by the US Army Research Office under Grant 
No.~W911NF-24-2-0228. 

\vspace*{0.1in}

\section*{Author Contributions}
\vspace*{-0.1in}
S.P., L.-W. K., B.G., M.H., and Y.-C. L. designed the research project.
S.P. and L.-W. K. performed the computations. All analyzed the data 
and results. S.P., L.-W. K. and Y.-C. L. wrote and edited the manuscript.

\vspace*{0.2in}
\section*{Competing Interests}
\vspace*{-0.1in}
The authors declare no competing interests.

\end{document}